\documentstyle[preprint,aps]{revtex}
\tightenlines
\begin{document}
\draft

\title{
Shake-up effects and intermolecular tunneling in C$_{60}$ ions.}

\author{F. Guinea \\}
\address{
	Instituto de Ciencia de Materiales.
	Consejo Superior de Investigaciones Cient{\'\i}ficas.
	Cantoblanco. 28049 Madrid. Spain.}

\author{J. Gonz\'alez \\}
\address{
	Instituto de Estructura de la Materia.
	Consejo Superior de Investigaciones Cient{\'\i}ficas.
	Serrano 123, 28006 Madrid. Spain.}

\author{M. A. H. Vozmediano \\}
\address{
	Escuela Polit\'ecnica Superior.
	Universidad Carlos III.
	Butarque 15.
	Legan\'es. 28913 Madrid. Spain.}

\date{\today}
\maketitle
\rightline{\it cond-mat/9403083}
\begin{abstract}

	The multiplet structure induced
by the Coulomb interactions in
C$_{60}^{n-}$ ($n = 1 - 12$) is analyzed. The partially ocuppied LUMO
gives rise to a set of levels which fill a relatively wide band
( $\approx 2-3$ eV ). A large ($\sim 1$ eV) intramolecular
effective repulsion is found. The anions are also shown
to be highly polarizable.
The optical absorption and photoemission
spectra are calculated. The
probability of exciting the
molecule upon the addition or removal of one electron
is shown to be significant, and this effect may give rise
to the insulating behavior of K$_4$C$_{60}$.
The role of intermolecular interactions in suppressing
the net repulsion within the C$_{60}$ molecules is discussed.

\end{abstract}


\narrowtext

	The small size of the C$_{60}$ molecule implies that
charging effects are significant. Estimates for the repulsion
between two additional electrons in a free
molecule are in the range of 3-4
eV\cite{Antropov,TDLee,PRB}. In a crystal, screening is expected
to reduce this value, to about 1 eV\cite{Antropov}.

	The most straightforward consequence of a
significant Coulomb repulsion is the existence of
different multiplets, when the outer shells are
partially filled\cite{multiplet1,multiplet2,multiplet3}.
In C$_{60}$, the lowest unoccuppied
molecular orbitals (LUMO's) are two triply
degenerate levels ($t_u$ and $t_g$), which lie close in energy
($\approx 0.5$ eV). Given the magnitude of the
Coulomb interactions, both these levels should
be considered when discussing the low energy multiplet structure.

	The excitations within these multiplets
are the likely cause of a feature identified
as a 'LUMO plasmon' in HREED experiments\cite{plasmon}.
They should also show up in optical absorption
measurements. There is a finite probability
that the injection, or emission of one electron
will leave the molecule in an excited state. These
shake-up processes give an intrinsic width to the LUMO peaks,
as measured in photoemission experiments.

	In a crystal, the intermolecular hybridization
is an additional source of broadening. Intramolecular
shake-up, and intermolecular tunneling are almost
incompatible processes.
Strong shake-up effects reduce intermolecular hopping, while
a large hybridization implies the irrelevance of
intramolecular interactions. Experimentally, a
significant width has been
observed\cite{photo1,photo2,photo3,photo4,photo5,photo6}.
Shake-up effects have also been discused in connection with
electron emission from a single C$_{60}$ molecule\cite{emiss}.

	The competition between shake-up and tunneling
has also consequences for other electronic
properties, like the conductivity. A large energy gap
between succesive multiplets, and a reduced hopping rate
favors insulating behavior, while the opposite happens if
tunneling dominates\cite{hubbard}.

	In the present work, we analyze the multiplet
structure of C$_{60}^{n-}$. We diagonalize the Coulomb
interaction in the space of the two LUMO's. We use a
continuum approximation, which correctly describes
these lowest lying orbitals\cite{PRL}.

	Within a continuum description, it can be shown that
the two lowest triplets correspond to the zero modes
of an effective Dirac equation\cite{PRL}. In this
scheme, the angular momentum is a good quantum number.
These states correspond to $l = 1$. The
wavefunctions in these two triplets
are\cite{PRB}:

\begin{equation}
\begin{array}{rlrl}
\Psi_{+1,a}^{\alpha} &= \sqrt{3 \over{8 \pi}} {\sin}^2 ( {\theta \over 2 } )
e^{i \phi} &\Psi_{+1,a}^{\beta} &= 0 \\
\Psi_{0,a}^{\alpha} &= \sqrt{3 \over{4 \pi}} \sin ( {\theta \over 2} )
\cos ( {\theta \over 2} ) &\Psi_{0,a}^{\beta} &= 0 \\
\Psi_{-1,a}^{\alpha} &= \sqrt{3 \over{8 \pi}} {\cos}^2 ( {\theta \over 2 } )
e^{ - i \phi} &\Psi_{-1,a}^{\beta} &= 0 \\
\Psi_{+1,b}^{\alpha} &= 0 \qquad\qquad
&\Psi_{+1,b}^{\beta} &= \sqrt{3 \over{8 \pi}} {\cos}^2
( {\theta \over 2} ) e^{i \phi} \\
\Psi_{0,b}^{\alpha} &= 0 \qquad\qquad
&\Psi_{0,b}^{\beta} &= - \sqrt{3 \over{4 \pi}}
\sin ( {\theta \over 2} ) \cos ( {\theta \over 2} ) \\
\Psi_{-1,b}^{\alpha} &= 0 \qquad\qquad
&\Psi_{-1,b}^{\beta} &= \sqrt{3 \over{8 \pi}} {\sin}^2
( {\theta \over 2} ) e^{- i \phi} \\
\end{array}
\end{equation}

	Each wavefunction is a spinor (components
labelled as ($\alpha , \beta$), reflecting the fact
that they are defined on a lattice with two sites per unit cell.
The subscripts $a,b$ refer to the two possible
'flavors' of the states.
This additional degeneracy (discussed in detail in\cite{TDLee,PRB})
is related to the well known 'fermion doubling problem'.
These levels,
which are degenerate and lie at zero energy
in the continuum approximation, are shifted and
split by the icosahedral symmetry of the molecule. The effect
can be included, in a phenomenological way, by introducing
a one body hamiltonian which mixes the two different
flavors. Within this scheme, the system is described
by means of two parameters only: the splitting already
mentioned, $= 2 \Delta$, and the magnitude of the
Coulomb repulsion, $E_C = e^2 / ( \epsilon R )$.

	Within the continuum description, the total angular
momentum, $L$, is a good quantum number, as is the total spin, $S$,
and the parity.
The existence of the flavor degeneracy
allows us to define an isospin operator, $S'$.

	The interaction between two electrons was discussed in
ref.\cite{PRB}.
The electron-electron term
of the hamiltonian, when projected onto the basis spanned
by the two lowest triplets (eq. (1)) can be written as:

\begin{equation}
{\cal H}_{e-e} = E_0 \big( k_1 \sum_{i,j} {\hat{n}}_i {\hat{n}}_j
+ k_2 \sum_{i,j} {\hat{\vec{l}}}_i {\hat{\vec{l}}}_j
+ k_3 \sum_{i,j} {\hat{\vec{s}}}_i {\hat{\vec{s}}}_j
+ k_3 \sum_{i,j} {\hat{\vec{s'}}}_i {\hat{\vec{s'}}}_j
+ k_4 \sum_{i \in a,j \in b} {\hat{\vec{l}}}_i {\hat{\vec{l}}}_j \big)
\end{equation}

	where $\hat{\vec{l}}$ is the
angular momentum operator, $\hat{\vec{s}}$
the spin,  and $\hat{\vec{s'}}$ the isospin.
 The energy scale $E_0$, and
the dimensionless constants $k_1 , k_2, k_3$ and $k_4$
depend on the nature of the interaction. For a simple
Coulomb repulsion, we have $E_0 = E_C, k_1 = 49/50,
k_2 = 11/50, k_3 = -3/25 $
and $k_4 = -1/2$. Other
interactions, such as a contact coupling,
modify the overall scale, but leave the relative
values of the $k$'s almost unchanged.
Equation (2)
makes explicit how exchange effects favor a high spin or
isospin.

	The local field splitting can be added to (2), as:

\begin{equation}
{\cal H}_{split} =  \Delta \sum_{m,s} {c^+}_{a,m,s} c_{b,m,s} + h. c.
\end{equation}

	Indices $a,b$ stand for the isospin, $s = \pm$ is the
real spin, and $m = 0, \pm 1$ is the angular momentum.

	In table I we give the lowest lying multiplets, and
the total width of the spectrum for $\Delta / E_C = 0.3$,
and $E_C = 1$eV. These parameters have been adjusted so as
to reproduce the ordering of the lowest levels, and the overall scale of the
multiplet structure of the discrete C$_{60}^{n-}$ anions, for
$n = 1 - 5$\cite{unpubl}, when the electrostatic interaction
is modelled to fit well the optical
absorption of neutral C$_{60}$\cite{absorption}.
The electron-electron repulsion in\cite{absorption}
is of the form: $V ( r ) = 1 / \sqrt{ 1 / U^2 + r^2 / ( r_0 V )^2 }$, where
$r_0$ is the average interatomic distance, $U = 7.2$ eV and
$V = 3.6$ eV\cite{mfa}.

	Some of the salient features of these results are:

	- There is always a net repulsion between electrons,
defined as $U_{eff} = E_{N-1} + E_{N+1} - 2 E_N$. Its is of order
$E_C$. It is somewhat reduced by screening induced
by rearrangements within the LUMO's. $U_{eff}$ shows local
maxima for $N = 3,6,$ and 9. If no other effects are
present, a crystal built up of C$_{60}^{3-}$ anions should be
an insulator\cite{photo1,hubbard}. Electrochemical
measurements\cite{electro} also suggest the existence of
this intramolecular repulsion.

	- While Hund's rule is not strictly followed, some of the lowest
lying states have high spins. These results may help to explain
the anomalously large signal found in ESR experiments\cite{ESR}.
The higher lying states tend to have lower spins, which is
also consistent with the observed decrease in the spin relaxation
rate with increasing temperature.

	- Multiplets tend to appear in pairs
with the same quantum numbers but opposite parity. This is related to
the approximate isospin symmetry described earlier.

	The electrical polarizability associated to transitions
within these multiplets is also listed in table I. These values
should be added to the polarizability arising from the remaining
$\pi$ levels. Within the same model\cite{ren}, this value is
90\AA$^3$.

	We have also analyzed the leading dipole allowed
transitions out of the C$_{60}^{2-}$, C$_{60}^{3-}$ and
C$_{60}^{4-}$ lowest states, which are given in table II.
In the convention that we use to define
the dipolar strengths, they  obey the sum rule:
$\sum_j | < i | \hat{P} | j > |^2 \le N$.

	The spectral strengths show a peak for
energies $\sim 0.6 - 0.8 E_C$, in all three cases.
Identification of this peak with the lowest feature
observed in\cite{plasmon},
implies that $E_C \sim 1$eV in K$_3$C$_{60}$, showing the adequacy
of the parameters employed.

	The previous analysis shows the existence of a large
intramolecular Coulomb repulsion, $\sim 1$ eV, a result consistent
with most theoretical analyses. Its scale is beyond the range
which can be reasonably compensated by intramolecular phonon effects.
Our results, however, also indicate the existence of a large
intermolecular polarizability. The bulk dielectric constant is
$\epsilon = 1 + 4 \pi \rho P$, where $\rho$ is the density of
C$_{60}$ molecules, and $P$ is the polarizability.
We find $\epsilon = 3.75$.  The reduction in energy
due to the polarization of a neutral molecule in the presence
of a unit charge at distance $D$ is: $\Delta E = e^2 / D ( P / D^3 )$.
Setting $D = 3 \times R$, where $R = 3.5$ \AA is the radius of
the C$_{60}$ molecule, we find $\Delta E = 0.18$ eV.
If we consider a charged molecule and its twelve nearest
neighbors, including all electrostatic interactions
between them, we find a polarization energy
$\Delta E \approx 1.6$ eV. Thus, the intermolecular
polarization can well compensate the intramolecular
repulsion. Other couplings can also contribute
to give a net intramolecular attraction, but, unless a very
precise cancellation takes place, the intermolecular
electrostatic interaction is the only one with the
right order of magnitude.
Note that no intermolecular interaction can influence
the multiplet splitting within a
given charge state.

	We have calculated the probability of exciting
the molecule upon the addition, or removal of a single electron.
The probability for transitions between different ground states
is 0.87 for the C$_{60}^{2-} \leftrightarrow$ C$_{60}^{3-}$ transition,
0.37 for the C$_{60}^{3-} \leftrightarrow$ C$_{60}^{4-}$ transition,
and 0 for the C$_{60}^{4-} \leftrightarrow $C$_{60}^{5-}$ one.
These numbers reduce intermolecular hopping. In the last case,
the lowest lying allowed transition is from
the ground state of C$_{60}^{4-}$ ( $^3$P$_+$ ) to the second excited
state of C$_{60}^{5-}$ ( $^4$D$_+$ ), which lies 0.38 eV above the
ground state (see table I). This mismatch may help to explain
the insulating behavior of K$_4$C$_{60}$. Metallic behavior in this
compound implies that the C$_{60}$ molecules fluctuate,
at least, between the 3$^-$, 4$^-$ and 5$^-$ charge states. If the
C$_{60}^{3-} \leftrightarrow$ C$_{60}^{4-}$ transition is
inhibited ( by a factor 0.37), and the
C$_{60}^{4-} \leftrightarrow$ C$_{60}^{5-}$ one is
completely suppressed, the system should be insulating.

	The coherence length in superconducting
M$_3$C$_{60}$, which gives the size of the Cooper
pairs,  is comparable to the C$_{60}$ radius. Thus, the creation of
a Cooper pair can be roughly described as a transition from
C$_{60}^{2-}$ to C$_{60}^{4-}$. The operator which
relates these two ground states is of the $^1$S type, which
should give the symmetry of the order parameter.

	In figure 1 we show the photoemission spectra of C$_{60}^{2-}$
C$_{60}^{3-}$ and C$_{60}^{4-}$. They have a non negligible width, although
the highest lying multiplets do not contribute. The existence of
two closely spaced levels with different symmetries in C$_{60}^{2-}$
gives a strong temperature dependence to the spectra, in agreement
with experimental results\cite{photo3}. These results suggest that
the width observed in photoemission is intrinsic to the charged
C$_{60}$ anion. It would be interesting
if similar features are observed in C$_{60}$ with metal atoms
in an endohedral position\cite{endo}.

	It is worth noting that the ground state of
most anions considered show orbital degeneracy (L = 1).
It will give rise to a Berry phase when quantizing the
intramolecular vibrations\cite{PRL,tosatti}. In addition,
the adiabatic transport of a charge also generates
a Berry phase. For instance, the motion of a charged
'vacancy' ( C$_{60}^{2-}$ ) in a background of 'neutral'
C$_{60}^{3-}$ leads to a non trivial phase. It will manifest
itself as a 'spin-orbit' coupling, where the 'spin' refers
to the angular momentum of the C$_{60}^{2-}$ anion,
and the 'orbit' is given by its motion throughout the lattice.
The effects of these phases will be studied elsewhere.

	The electrostatic interactions considered here
do not rule out the relevance of couplings
mediated by phonons\cite{tosatti}.
Ihe intramolecular charging energy depends on the radius
of the C$_{60}$ molecule, which is modified by breathing modes.
More importantly, changes in the intermolecular distances
alter the polarization energy, which we are assuming to give
the leading contribution to a net attraction. The main physical
effects should be approximately described
by an attractive Hubbard model at half filling,
described in terms of the intermolecular hopping,
$t \sim 0.02 - 0.04$ eV\cite{hopping}, and the intramolecular
attraction, $| U | \sim 0.1 - 0.3$ eV. This model is a
superconductor, with $T_c \sim | U | e^{- W / | U |}$,
where $W = 16 t$ is the bandwidth. Thus, we expect a dependence
of $T_c$ on the lattice parameter, and the existence
of an isotope effect. The dependence of $U$ on the mass
of the carbon atoms is not the conventional one, $\sim M^{-1/2}$,
and the isotope effect needs not show standard behavior.

	The previous study shows that electrostatic
interactions dominate over other electron-electron couplings.
Intramolecular charging effects should give rise to insulating
behavior, but intermolecular electrostatic interactions can compensate
this tendency. We have analyzed the influence of the electrostatic
interactions on properties
which lie beyond the scope of usual band structure calculations,
like the satellite structure in photoemission and the renormalization
of intermolecular hopping.
The global picture obtained here is consistent with the observed linewidth
in photoemission spectra and their temperature dependence. It also
shows a 'LUMO plasmon' in good correspondence with HREED experiments.
The shake-up processes upon intermolecular charge transfer
give a reasonable explanation for the insulating behavior in
K$_4$C$_{60}$.

\newpage
\begin{table}

\caption{Lowest lying multiplets for charged states of C$_{60}$.
The parameters used are: $E_C = 1$eV, $\Delta = 0.3$eV.
States are classified in terms of their spin, angular momentum
and parity. Energies are in eV. The ground state polarizability
(in \AA$^3$) is also listed.}

\begin{tabular}{l|c|l|c|l|c}
\multicolumn{2}{c|}{C$_{60}^{-}$} &
\multicolumn{2}{c|}{C$_{60}^{2-}$} &
\multicolumn{2}{c}{C$_{60}^{3-}$} \\
\hline
\multicolumn{2}{c|}{Total width 0.600 } &
\multicolumn{2}{c|}{Total width 1.581 } &
\multicolumn{2}{c}{Total width 2.585 } \\
\multicolumn{2}{c|}{Polarizability 73.9 } &
\multicolumn{2}{c|}{Polarizability 67.6 } &
\multicolumn{2}{c}{Polarizability 39.6 }\\
\tableline
state & energy & state & energy & state & energy \\
\hline
	 $^2$P$_+$ & -0.300 &
	 $^3$P$_+$ &  0.300 &
         $^4$S$_-$ &  1.743 \\

	 $^2$P$_-$ &  0.300 &
	 $^1$S$_+$ &  0.319 &
         $^2$D$_-$ &  1.912 \\

     & & $^1$D$_+$ &  0.360 &
         $^2$P$_-$ &  1.931 \\

     & & $^1$S$_-$ &  0.600 &
         $^4$S$_+$ &  1.946 \\

     & & $^3$P$_-$ &  0.700 &
         $^4$D$_+$ &  2.140 \\

     & & $^3$D$_-$ &  0.760 &
         $^2$P$_+$ &  2.252 \\

\end{tabular}
\begin{tabular}{l|c|l|c|l|c}
\multicolumn{2}{c|}{C$_{60}^{4-}$} &
\multicolumn{2}{c|}{C$_{60}^{5-}$} &
\multicolumn{2}{c}{C$_{60}^{6-}$} \\
\hline
\multicolumn{2}{c|}{Total width 4.009 } &
\multicolumn{2}{c|}{Total width 4.542 } &
\multicolumn{2}{c} {Total width 5.137 } \\
\multicolumn{2}{c|}{Polarizability 48.8 } &
\multicolumn{2}{c|}{Polarizability 67.5 } &
\multicolumn{2}{c} {Polarizability 61.1 } \\
\tableline
state & energy & state & energy & state & energy \\
\hline
	 $^3$P$_+$ &  4.383 &
	 $^2$P$_-$ &  7.941 &
	 $^1$S$_+$ & 12.419 \\

	 $^1$D$_+$ &  4.476 &
	 $^2$P$_+$ &  8.254 &
	 $^1$S$_-$ & 12.671 \\

	 $^1$S$_+$ &  4.486 &
	 $^4$D$_+$ &  8.323 &
	 $^3$D$_-$ & 12.902 \\

	 $^3$P$_-$ &  4.664 &
	 $^4$P$_+$ &  8.357 &
	 $^3$S$_-$ & 12.950 \\

	 $^5$P$_-$ &  4.700 &
	 $^4$S$_+$ &  8.369 &
	 $^3$P$_-$ & 12.960 \\

	 $^1$S$_-$ &  4.719 &
	 $^2$F$_+$ &  8.454 &
	 $^1$S$_+$ & 12.981 \\
\end{tabular}
\end{table}

\begin{table}
\caption{Leading dipole allowed transitions from low lying multiplets
for different charge states of C$_{60}$. The parameters are the same as
in table I.}

\begin{tabular}{c|l|l|l}
Charge state & transition & energy (eV) & dipolar strength \\
\tableline
C$_{60}^{2-}$ &
	 $^3$P$_+  \rightarrow  ^3$D$_-$ & 0.460 & 0.410 \\
&	 $^3$P$_+  \rightarrow  ^3$S$_-$ & 1.300 & 0.821 \\
&	 $^1$S$_+  \rightarrow  ^1$P$_-$ & 1.121 & 0.480 \\
&	 $^1$D$_+  \rightarrow  ^1$P$_-$ & 1.080 & 0.923 \\
&	 $^3$D$_-  \rightarrow  ^3$P$_+$ & 0.840 & 0.923 \\
\hline
C$_{60}^{3-}$ &
	 $^4$S$_-  \rightarrow  ^4$P$_+$ & 0.957 & 0.854 \\
&	 $^2$D$_-  \rightarrow  ^2$F$_+$ & 0.568 & 0.262 \\
&	 $^2$D$_-  \rightarrow  ^2$P$_+$ & 0.604 & 0.127 \\
&	 $^2$D$_-  \rightarrow  ^2$P$_+$ & 1.390 & 1.061 \\
&	 $^2$P$_-  \rightarrow  ^2$D$_+$ & 0.483 & 0.280 \\
\hline
C$_{60}^{4-}$ &
	 $^3$P$_+  \rightarrow  ^3$D$_-$ & 0.840 & 0.526 \\
&	 $^3$P$_+  \rightarrow  ^3$S$_-$ & 1.137 & 0.526 \\
&	 $^1$D$_+  \rightarrow  ^1$F$_-$ & 0.803 & 0.431 \\
&	 $^1$D$_+  \rightarrow  ^1$P$_-$ & 1.602 & 1.382 \\
&	 $^1$S$_+  \rightarrow  ^1$P$_-$ & 1.002 & 0.922 \\
\end{tabular}
\end{table}

\figure{ Figure 1. Direct (full lines) and inverse (dashed lines)
photoemission spectra of different
C$_{60}^{n-}$ anions. The temperature dependence in the spectra
of C$_{60}^{3-}$ and C$_{60}^{4-}$ is negligible.}

\end{document}